\documentclass[twocolumn,superscriptaddress,secnumarabic,amssymb,nobibnotes,aps,prl]{revtex4-2}

\usepackage{extarrows}
\usepackage{amsmath}    % need for subequations
\usepackage{graphicx}   % need for figures
\usepackage{verbatim}   % useful for program listingsz
\usepackage{color}      % use if color is used in text
\usepackage{subfigure}  % use for side-by-side figures
\usepackage[colorlinks=true, allcolors=blue]{hyperref}   % use for hypertext links, including those to external documents 
\usepackage{verbatim}   % and URLs
\usepackage{longtable}
\usepackage{epsfig}
\usepackage{comment}
\usepackage{bm}
\usepackage{ulem}
\usepackage{dcolumn} % Align table columns on decimal point
\usepackage{textcomp}   % use \textdegree
\usepackage{kotex}      % use Korean
\usepackage[version=3]{mhchem}  % to use this : \ce{}
\usepackage{xcolor}

\newcommand\redout{\bgroup\markoverwith
  {\textcolor{red}{\rule[0.5ex]{2pt}{0.8pt}}}\ULon}

\newcommand{\lnof}{\ce{La3Ni2O5F}}

\begin{document}

\title{Dichotomous electronic system in a bilayer Ni$^{1+}$ nickelate}
\author{Young-Joon Song}
\affiliation{Institut f\"ur Theoretische Physik, Goethe-Universit\"at Frankfurt, Max-von-Laue-Stra\ss e 1, 60438 Frankfurt am Main, Germany}
\affiliation{Asia Pacific Center for Theoretical Physics, Pohang, 37673, Korea}
\author{Kwan-Woo Lee}
\email{mckwan@korea.ac.kr}
\affiliation{Division of Semiconductor Physics, Korea University, Sejong 30019, Korea}
\affiliation{Department of Applied Physics, Graduate School, Korea University, Sejong 30019, Korea}
\author{Warren E. Pickett}
\email{wepickett@ucdavis.edu}
\affiliation{Department of Physics and Astronomy, University of California, Davis, California 95616, USA }
\date{\today}
%\pacs{}

\begin{abstract}
``Infinite layer'' nickelates ${\cal R}$NiO$_2$ (${\cal R}$=rare earth), having empty apical O sites, become superconducting upon hole doping, stimulating research into the related sequence Nd$_{n+1}$Ni$^{+p}_n$O$_{2n+2}$, formal charge state $p$=1+$\frac{1}{n}$, $n$=2,3,4,...., with the $n$=5 member being found to be superconducting. The two layer system La$_3$Ni$_2$O$_{7-\delta}$, with $\delta$=0,$\frac{1}{2}$,1 ($p$=2.5,2,1.5) approaches the peak in the nickelate superconducting dome but shows no superconductivity.  
Newly reported \lnof~reaches the  Ni$^{1+}$ goal while, as we show, introducing a partially occupied {\it electron} band $E^*$, based on an interstitial density that extends over the three open ``apical'' layers and leads to a single cylindrical electron Fermi surface giving self-doping. The commonly inert Ni $d_{xz},d_{yz}$ orbitals partner with interstitial $E^*$ to provide an incipient non-analytic Dirac point, with the critical point being reachable by pressure or further F insertion. The $E^*$ electron cylinder and the conventional Ni $dp\sigma$ hole carriers combine to provide a two-fluid dichotomy of hole and electron quasiparticles, affecting normal state properties that should verify the dichotomous aspect of transport. 
\end{abstract}
\maketitle
\clearpage

{\it Background.}
Electride crystals entered the lexicon of solid state physics and chemistry as an interstitial density, containing one electron and not ascribable to any atomic orbital: an `isolated' electron acting as an anion $e^{-}$. These occurred in molecular crystals, where empty spaces could provide a suitable potential, and the available space was sufficient to obviate a positive energy from confinement. An example that began the excitement, and potential for novel properties, was Dye's discovery \cite{dye1993,wagner1994}, and a subsequent calculation \cite{djsingh1993}, that verified the interstitial density. Two crown ether molecules (15-crown-5) encased Cs, expelling its weakly bound valence electron into a space between the roughly spherical pair of crown ether molecules -- the Cs$^+$(1-crown-5)$_2$:$e^-$ electride. This crystal provided a singly occupied (magnetic) electride gap state \cite{djsingh1993}, closely related to color center defects. In the intervening decades several review articles have related how molecular electrides display unexpected properties: electrical conduction, magnetism (namely, the localized electron mentioned above), surface catalytic activity, and even superconductivity (SC) \cite{c.liu2020,hosono2021,f.sun2026}. 
In the meantime, interest has turned more toward inorganic electrides, with application of the term `electride' having undergone generalization, being adopted whenever there is an unexpected density maximum not ascribable to an atomic or molecular orbital. Alkali metals under pressure have assumed unanticipated structures with `electride' regions \cite{w.cheng2022,alkalis}. At layered van der Waals interfaces an interstitial density can arise \cite{hy.song2021,j.zhou2024}. The advent of rapid computational methods has led to a rise in predicted inorganic electrides, often at high pressure.
For example, it is predicted that at high pressures of the Earth inner core,
electride formation can stabilize the hcp iron lattice \cite{dy.kim2024}.

%%%%% Fig1

\begin{figure}[h!]
\centering
\includegraphics[width=0.8\columnwidth]{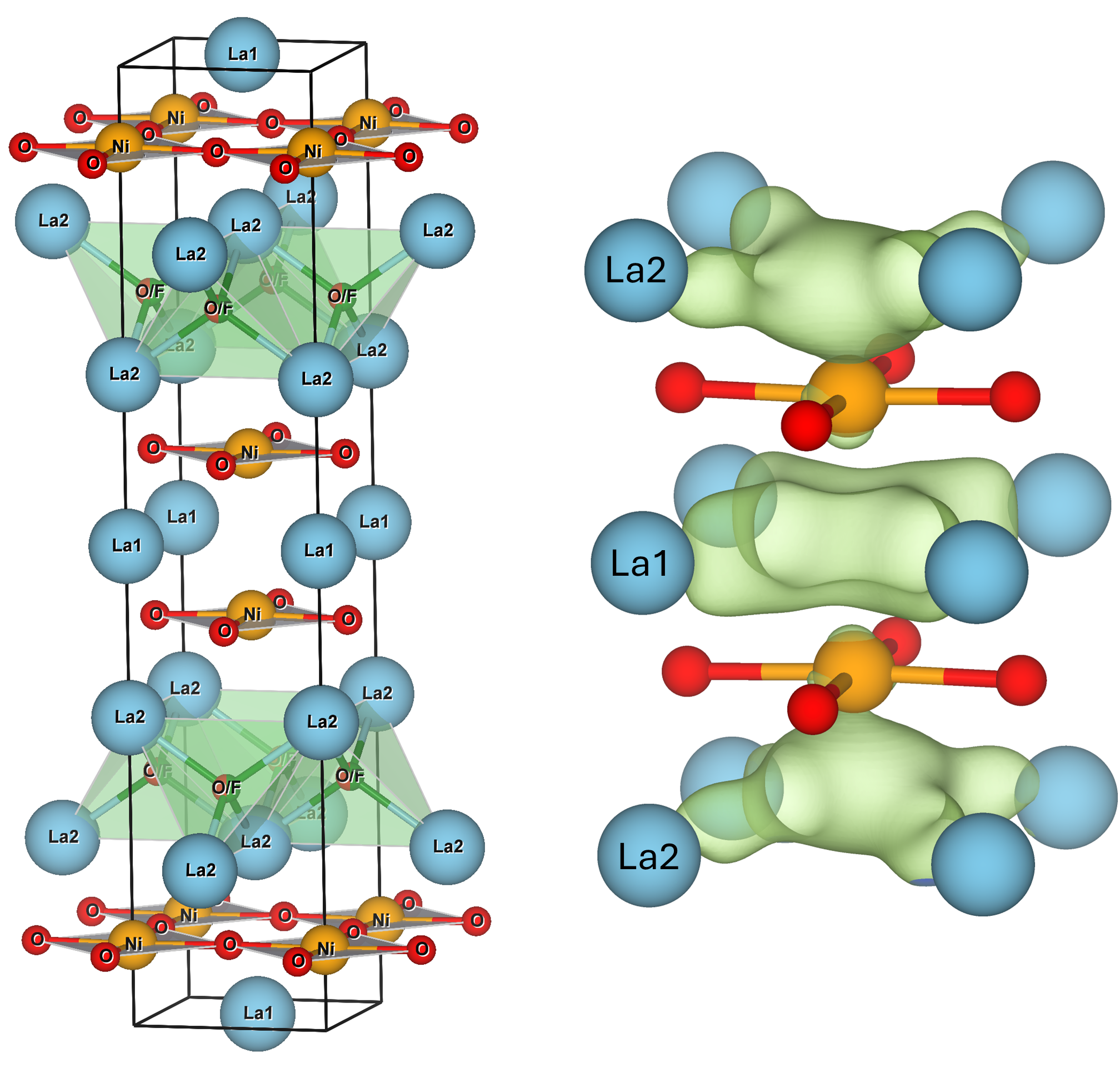}
\caption{Left: Tetragonal $I4/mmm$ structure of \lnof, consisting of double NiO$_2$ infinite layers isolated by the La2 and La1 layers. Right: Isosurface plot of the occupied $E^*$ wavefunction $|\Psi(r)|$ at $M$ (see text). The $E^*$ density, arising from a single band, consists of roughly equal parts in each of the three La-apical site layers, with connections (see text).}
\label{fig:str}
\end{figure}  

%%%%%%% FIG 2
\begin{figure*}[ht]
\centering
\includegraphics[width=1.99\columnwidth]{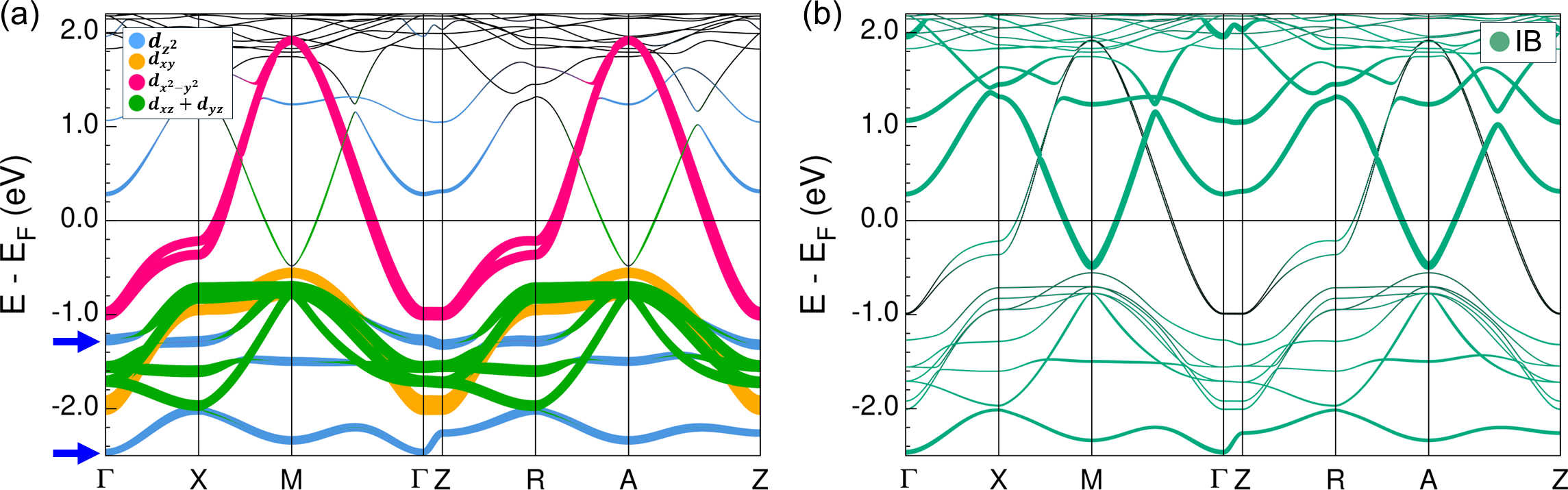}
\caption{Nonmagnetic band structure of \lnof. (a) Ni $3d$ orbital decompositions are distinguished by color.  The blue arrows indicate the large splitting of the $d_{z^2}$ orbitals at the $\Gamma$ point due to coupling with, or through, the interstitial ($I$) density in the intervening La1 layer. Points are as follows: Except for the usual $dp\sigma$ band, the other Ni $d$ bands lie at $-1.3\pm$ 0.5 eV, hence are electrically inactive; $t_{2g}$ splitting is small.
(b) Fatband representation of the $I$ character, that is, the amplitude of the state not falling within any atomic sphere. The $E^*$ band dips down from 2 eV at $\Gamma$ to $-0.5$ eV at $M$. See the text for other $I$ density-related features.
}
\label{fig:nm_fat}
\end{figure*}

These uses of `electride' have been applied where the amount of localized charge is only a fraction of one electron, and to materials where the interstitial charge is delocalized. Their properties extend the use of `electride' to a much broader class of crystalline solids, and distinguishing them from materials with related properties can become unclear. So far relatively few proposed electrides have been realized in the laboratory \cite{c.liu2020,hosono2021,hy.song2021}.

{\it Nickelates.}
While at first glance appearing to be a distinct topic from the above, we extend the vigorous activity following the discovery {\cite{hwang2019}} of long-sought nickelate SC in hole-doped `infinite layer' nickelate (ILN) NdNiO$_2$, with formal Ni$^{p+}$ valence $p\approx$1.2. SC was next found in the $n$=5 member of the series Nd$_{n+1}$Ni$_n$O$_{2n+2}$ system \cite{pan2022}, formal valence $p$=$1+\frac{1}{n}$, with the same formal value $p$=1.2. This series is now one focus of attention, a particular point being that the $p$=1 (Ni$^{1+}$) formal charge state has been challenging to stabilize. 
Recently the $n$=4-7 systems have been found to be superconducting, providing another SC dome similar to that of hole-doped ILN, peaking at $p$=1.2  \cite{mundy2026}.

A related system is La$_3$Ni$_2$O$_{7-\delta}$, which for the varying O concentration $\delta$=0,$\frac{1}{2}$,1 gives $p$=2.5, 2, 1.5, respectively. La$_3$Ni$_2$O$_6$ \cite{greenblatt2006} has received the most attention, experimentally \cite{curro2013,egami2022,m.wang2023,cava2024} (it is insulating, possibly due to charge disproportionation), in the peculiarities of electronic structure \cite{pardo2011,botana2016,fp2022}, and study of possible SC \cite{dagotto2024,lechermann2025,j.hu2025,kuroki2026}. The properties of these compounds vary greatly and include an insulating regime, with no SC \cite{cava2024}. If the anion concentration could be reduced further, to $\delta$=1.5, the Ni$^{1+}$ charge state would be achieved. This goal has been reached with the announcement by Wernert {\it et al.} \cite{hayward2026} of La$_3$Ni$_2$O$_{5}$F=La$_3$(NiO$_2$)$_2$OF, replacement of one O ion in the blocking layer by F, thus the removal of $\frac{1}{2}$ additional hole per Ni, $p$=1.

From density functional theory studies we have found that \lnof~has an interstitial density that provides a different viewpoint on the behavior of, and independent insight into, interstitial density upon physical behavior. At the electronic carrier level, \lnof~provides two types of quasiparticles: one is the usual nickelate $dp\sigma$ Fermi surface (FS), 
and the other is essentially a planewave-like density confined to three channels. The spectrum of low-energy excitations (hence transport and far-IR properties) consists of each of these two distinctly different -- dichotomous -- quasiparticles, and their coupling via inter-FS scattering. 

{\it Results.}
Unexpectedly, we find that \lnof~contains a `blocking layer' seemingly typical of related cuprates and nickelates, but that results in
extreme blocking, indicated by all $3d$ bands of interest having negligible $k_z$ dispersion -- uncoupled bilayers.
The bottom of the two $dp\sigma$ bands (indistinguishable in plots) disperse by ~$\sim$0.5 meV, with similar (lack of) dispersion of the interstitial band (see below) at $-0.5$ eV at $M$ -- truly isolated electronic layers.

Representing the disordered O$_{0.5}$F$_{0.5}$ by the virtual crystal average X, the layer stacking sequence, shown in the left panel of Fig.~\ref{fig:str}, is \lnof~=[(La2)X(La2)][(NiO$_2$)(La1)(NiO$_2$)], i.e., a blocking layer and a NiO$_2$ bilayer.  
The virtual crystal treatment of O/F is justified because (i) O and F are neighbors in the periodic table, and (ii) the closed shell $2p$ states in this layer are strongly bound ($\sim$5-7 eV) and serve mainly to specify the electron doping of the NiO$_2$ layers. The formal Ni$^{1+}$ valence is isovalent to Cu$^{2+}$ but unusual, and very different in behavior \cite{LP2004,LP2020a,LP2020b,ours2026}. The methods of calculation, using {\sc wien2k} \cite{w2k}, are described in a separate study of the magnetic behavior of \lnof~\cite{ours2026}.

\begin{figure}[t]
\centering
\includegraphics[width=1.0\columnwidth]{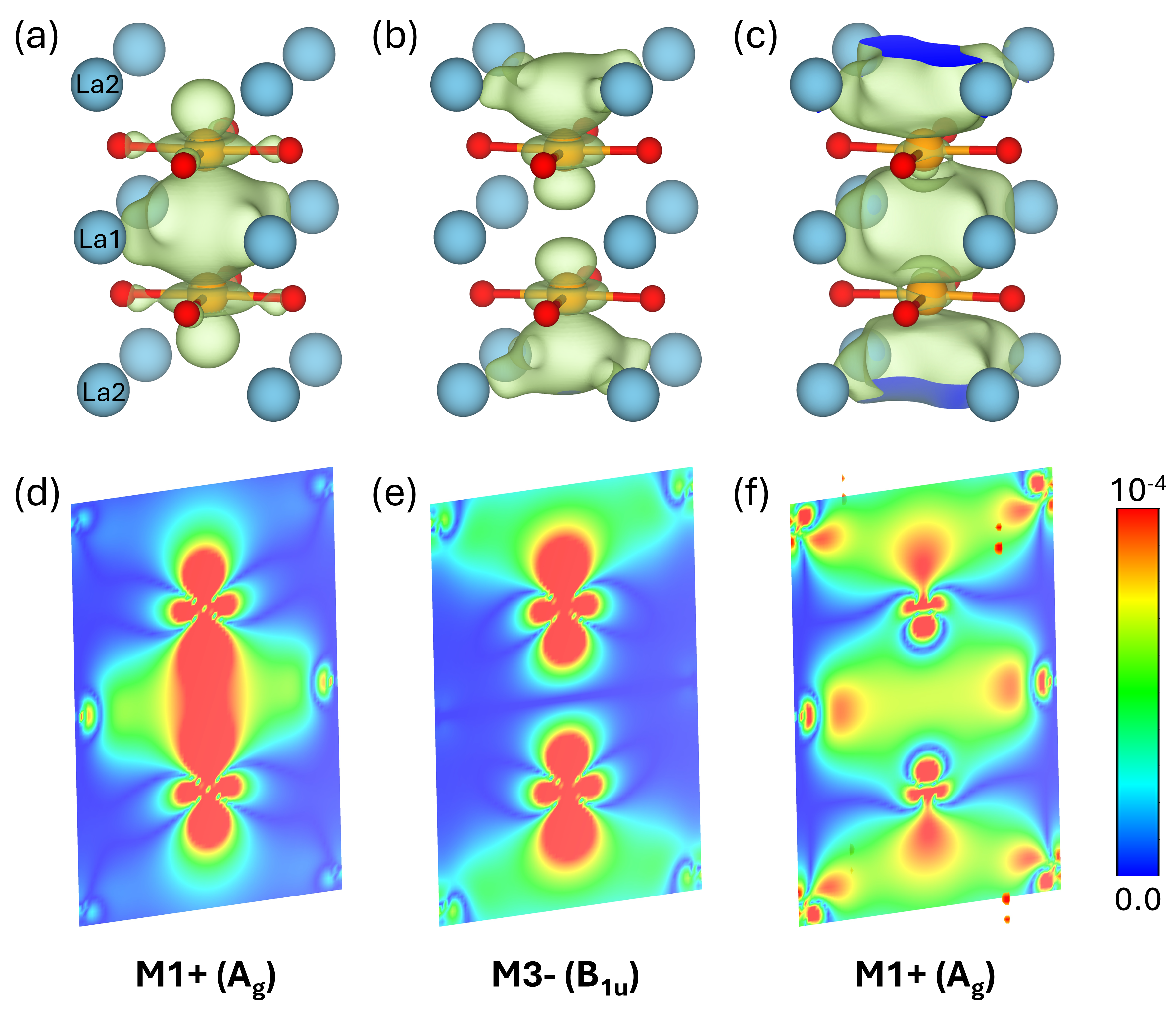}
\caption{Top row: Isosurface plots of wave functions $|\psi(\textbf{r})|$: the even [panel (a)] and odd [panel (b)] $d_{z^2}$ pairs at the $M$ point, at $-2.4$ eV and $-1.5$ eV respectively, and (c) the $E^*$ state at $-0.5$ eV. 
Bottom row [panels (d)-(f)]: Corresponding 2D-projected color-scale plots on the [110] plane, passing through La and Ni atoms, making the $z$ reflection symmetry clear, as well as the involvement of $I$ density.
Figure~\ref{fig:nm_fat} shows that the $d_{z^2}$ pair can be followed throughout the zone, with splitting of $\sim$1 eV. The $D^*$ state approaching $M$, part of the network as described in the text, is shown in Fig. S7 of the SM \cite{sm}.
}
\label{fig:M-wvfns}
\end{figure}

Some basic points related to the \lnof~bands and magnetic tendencies have been noted separately \cite{ours2026}, we elaborate here.  
In discussing the electronic structure, it is most instructive to begin above E$_F$ in the band structure, shown with orbital characters in Fig.~\ref{fig:nm_fat}, with more detail shown in the supplemental material (SM) \cite{sm}.\\
%%%%%%%%%%%%%%
$\bullet$ The usual focus, the pair of $dp\sigma$ bands, forms a twofold degenerate (except for some splitting near $X$) pair of large cylindrical hole FSs around $M$.
The FSs are displayed in Fig. S1 of SM \cite{sm}.\\
%%%%%%%%%%%%%%%%%%%
$\bullet$ A single ``$E^*$'' band dips from 1.5-2 eV above $E_F$ to $-0.5$ eV at $M$, with {\it linear} bands (except extremely near $M$) along both symmetry directions from $M$, up to +1 eV. This $E^*$ band derives from the $I$ density, emphasized in the right panel of Fig.~\ref{fig:str}. Being unassociated with any atom, this is essentially a 2D plane-wave-like band, confined by avoidance of the blocking layer and O$^{2-}$ ions. This band forms an electron cylinder around $M$ enclosing 0.18 electrons. As a result, the $dp\sigma$ bands are not half filled, with each Ni ion valence becoming +1.09.  This self-doping pushes \lnof~toward the SC dome of Ni$^{1+}$ nickelates \cite{d.li2020,s.zeng2020b,osada2020prm,pan2022} and the itinerant $E^*$ carriers can be expected to hinder magnetic ordering.\\
%%%%%%%%%%%%%%%%
$\bullet$ 
The $d_{z^2}$ orbitals, though lying 1--2 eV below $E_F$ with small dispersion, are even-odd split by $\sim$1 eV throughout the zone due to strong coupling to and through the $I$ density, see Fig. \ref{fig:nm_fat}. The coupling produces a lower even symmetry hybrid band $d_{z^2}^e$ with large $I$ density in the La1 layer apical site, and an upper odd partner $d_{z^2}^o$ with $I$ density maxima in the La2 layers, both pictured in Figs.~\ref{fig:M-wvfns}(a,b) and \ref{fig:M-wvfns}(d,e). \\
%%%%%%%%%%%%%%%%%%%%%%%
$\bullet$ The $I$ density avoids the $d_{x^2-y^2}$ and O $p_{\sigma}$ lobes, minimizing coupling between the $dp\sigma$ and $E^*$ bands, thereby producing a dichotomous electronic structure.\\ 
%%%%%%%%%%%%%%%%%%%%%
$\bullet$ Less evident is a shadow feature involving the $d_{xz}, d_{yz}$ bands lying 1--2 eV below $E_F$ with only 1 eV dispersion. Upon approaching $M$ from both symmetry directions, there is a $d_{xz}, d_{yz}$ band approaching the cluster of $t_{2g}$ bands at $M$, with the $E^*$ band just above at $-0.5$ eV. This ``$D^*$'' band is parallel to the $E^*$ band, with the same velocity and linearity.\\
%%%%%%%%%%%%%%
$\bullet$ This $D^*$ band acquires some $I$ density character upon nearing $M$; in complementary fashion, the $E^*$ wave function at $E_F$ acquires $d_{xz}, d_{yz}$ character creeping in. These features 
(wave-function plots are provided in Fig. S7 of the SM \cite{sm}) indicate the source of the $D^*$-$E^*$ pairing that gives an incipient Dirac point. 
Additional hole doping or compressive strain (see below) produces a nonanalytic critical Dirac point with unusually distinct, physically separated orbital character partners above and below the degeneracy. Moreover, this coupling provides the `glue' that cements the three layers of $I$ density into a single entity and band.\\
%%%%%%%%%%%%%

\vskip 2mm
 
 The $E^*$ band in \lnof~differs in a few ways from the electride band in ILNs besides the distinctive difference in the distribution of $I$ density. First, strong $k_z$ dispersion (2 eV) of the ILN electride band leaves only a pillbox FS centered on corner point $A$ versus the cylindrical $E^*$ FS along the zone corner $M$-$A$ line. Second, previous calculations on ILNs \cite{arita2019,nomura2022,adhikary,y.gu2020,sawatzky2023,j.you2026} have introduced a ``pseudo-atomic sphere'' with a zero nuclear charge `s' orbital in the interstitial region, enabling representation with maximally localized Wannier functions
including the `s' orbital. The interstitial
orbital corresponds to a spherical \cite{arita2019,adhikary,y.gu2020,nomura2022} or barrel-shaped \cite{sawatzky2023} shape, with details provided for NdNiO$_2$ \cite{sawatzky2023}. References \cite{arita2019,nomura2022} have provided densities of individual states.
Finally, our attempts to Wannierize including the $E^*$ band with a minimal basis set were unsuccessful.

It is common for bands several eV above E$_F$ to be interstitial planewave-like, avoiding atoms; it is very uncommon for such a band to dip down across E$_F$ and complicate the electronic structure as it does in \lnof, and in a linear band fashion.
The connection between $D^*$ and $E^*$ is seen by taking the La1 site as the center of $4/mmm$ symmetry. On the bottom Ni layer, rotating the Ni $d_{xz},d_{yz}$ around the plaquette successively by $\frac{\pi}{4}+\frac{m\pi}{2}$ ($m$=0,1,2,3), the linear combination is symmetric under 90$^{\circ}$ rotations. It is odd under $z$ reflection, so taking the odd combination with the upper Ni layer gives an orbital of $4/mmm$ symmetry, the same as $E^*$. Figure \ref{fig:M-wvfns}(c) shows that the $E^*$ wavefunction indeed has maxima in these directions toward La1 (and La2). One can form a $4/mmm$ symmetric state from $d_{z^2}$ functions using the even combination of lower and upper layers, essentially the $d_{z^2}^e$ band discussed above, but most of its spectral weight is expended in that low $d_{z^2}^e$ band.

\begin{figure*}[tb]
\centering
\includegraphics[width=2.0\columnwidth]{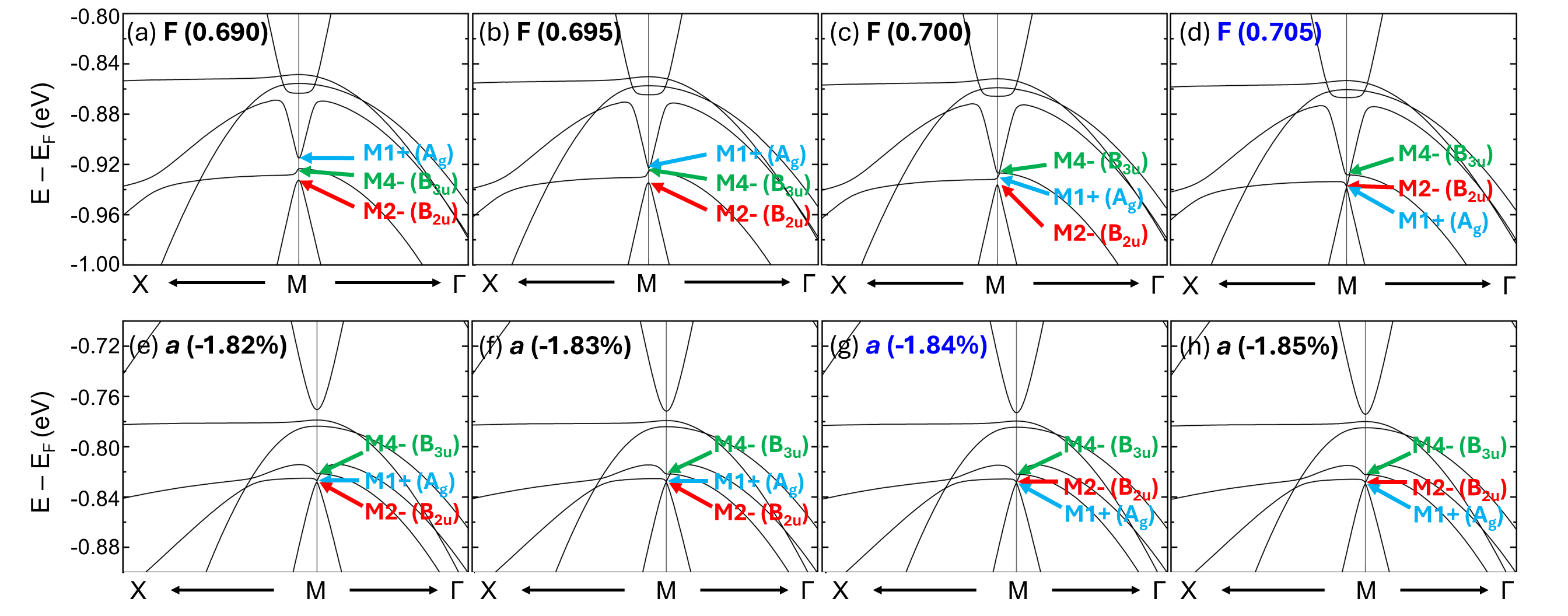}
\caption{Evolutions of the $E^*$ and $D^*$ bands, meeting and crossing, enlarged from the corresponding region of Fig.~\ref{fig:str}. 
Top row: Increasing the F concentration through the critical point at F(0.700) in La$_3$Ni$_2$O$_4$(O$_{1-x}$F$_x$)$_2$. Note the switching of eigenvalue order at F(0.705).
Bottom row: Applying symmetric strain, with a critical point at $\Delta a$=$-1.84$\%. The lattice parameter $a$ is changed as noted, with fixed volume. Decreasing $a$ could be achieved by pressure, since the change in $c$ may have negligible effect. In each case, the upper $E^*$ band shifts downward at $M$, meeting the $D^*$ band. 
}
\label{fig:F_a}
\end{figure*}   

{\it The Dirac point.} 
The $D^*$ and $E^*$ bands are anomalous and coupled as mentioned: (i) in approaching $M$ linearly,  (ii) doing so with parallel dispersions and equal velocities in both high symmetry directions, and 
(iii) thus they are evident continuations of each other across a small to vanishing gap. The origin is an incipient degeneracy and non-analyticity of the subeigensystem. There is only minor sensitivity of these bands to shifts of positions (internal parameters) of the Ni, La, and O atoms. A related  unoccupied band is evident in the study of the La$_{n+1}$Ni${_n}$O$_{2n+2}$ system with empty apical sites \cite{Antia2021,Antia2026}.

We have found sensitivity to the composition of the X virtual crystal atom: As the F composition of X=${\rm O/F}$ is increased from 0.5, the $E^*$ and $D^*$ bands at $M$ close up and touch at an F fraction near 0.700. This Dirac point is a tuned degeneracy protected by preservation of the crystal symmetry (see the top row of Fig. \ref{fig:F_a}) and it signals a nonanalytic eigensubsystem.  As the bands approach, the local band structure that includes other $d$ orbitals becomes intricate near the critical point.    As mentioned, the orientation of the $d_{xz}, d_{yz}$ orbitals enables mixing with the $E^*$ band (see the SM \cite{sm}), finally closing to the Dirac point.

Increasing the F fraction from 0.5 would be accomplished by extending the O$\rightarrow$F replacement process during synthesis. This replacement increases electron doping, evident in the lowering of the $E^*$ minimum at $M$ from $-0.5$ eV to $-1$ eV for F(0.7). 
Moving $E_F$ from half-filling will further lessen the tendency toward magnetic order, and move the Ni formal valence closer toward the peak in the SC dome. We have found that compressive strain, see the bottom row of Fig.~\ref{fig:F_a}, moves the system through the critical degeneracy with a somewhat different local band structure and band inversion. 

This $D^*$-$E^*$ Dirac point has characteristics in common with one in three-dimensional (3D) skutterudite CoSb$_3$ \cite{skutterPRL,skutterPRB}. In that compound, linear bands meet at $\Gamma$ and lie (with much import) at E$_F$. The upper partner is a triplet at $\Gamma$; only one linear combination of the triplet mixes with the lower linear band. The other point of interest is that the ``interstitial'' band does not have a density maximum in the large void in the skutterudite structure, instead it arises from a linear combination of Sb ``$4p_z$" orbitals perpendicular to the skutterudite `squares' surrounding the void. This void is large enough to hold an atom, hence the intense interest in `filled skutterudites' with unusual properties \cite{skutterSales}.

The interstitial density band in ILNs, discovered in 2004 \cite{LP2004} and for 15 years assigned to ``La $d$'' character until recently interpreted as ``electridelike'', has been little discussed in terms of its contribution to physical properties.
The dichotomous electronic structure of \lnof~involves 2D carriers of opposite sign, separated in both $k$ and real spaces, and with distinct characters -- a bilayer of correlated local $dp\sigma$ orbitals within layers of interstitial electrons confined to three channels. 

{\it Transport.}
Our separate study \cite{ours2026} indicates that the $dp\sigma$ FS will experience the magnetic fluctuations that are the likely dominant scattering, and pairing, mechanism, but that the $E^*$ FS is almost unaffected by Ni moments, confirmed by Madani {\it et al.} \cite{Madani2026} Our calculations of La and NiO$_2$ optical phonon coupling to the $E^*$ FS (see the SM \cite{sm}) give deformation potentials of $\sim$0.1 eV/\AA, an order of magnitude smaller than those in conventional phonon-coupled SCs. Moreover, Hausoel {\it et al.} \cite{Hausoel2025} have shown that dynamical electron-electron interactions in ILNs leave the corresponding band in Nd$_{n+1}$Ni$_n$O$_{2n+2}$ compounds ($n$=4-7) unaffected except for band filling due to renormalization of the $dp\sigma$ band. The $E^*$ band in \lnof~thus seems almost immune to disturbances.

Based on this evidence, we propose a two-band picture with very different values of scattering time $\tau$ for the two carriers. The factor of 5 larger value of $\Omega_{P,dp\sigma}^2$ for the $dp\sigma$ bands will therefore be compensated (perhaps overcompensated) by the smaller value of $\tau_{dp\sigma}$, so the $E^*$ FS may contribute strongly to the conductivity, somewhat short-circuiting the resistivity to values lower than those in other nickelates. 

Within Bloch-Boltzmann theory, the conductivity is the sum of contributions from each FS (neglecting the very small difference in velocities)
\begin{eqnarray}
 \sigma_{xx}(T)&=&
 \frac{2}{4\pi}[ \Omega_{P,dp\sigma}^2\tau_d(T)
   +\Omega_{P,E}^2\tau_E(T) ]\nonumber \\
   &\rightarrow& \frac{e^2 v}{h}
               [\tau_p (2\pi n)^{1/2}      
        +       \tau_n (2\pi p)^{1/2}]),
\label{eqn:conductivity}
\end{eqnarray} 
where the factor of 2 is for the square lattice case, and $\sigma_{yy}=\sigma_{xx}$ by symmetry. The simple band structure (circular, isotropic) allows the FS quantities to be expressed in terms of the hole and electron densities $p,n$ and $\tau_p,\tau_n$ ($dp\sigma$, $E^*$, respectively). Their separate FS density of states, Fermi velocities, and Drude plasma frequencies $\Omega_{P}$ are given, respectively, ($dp\sigma$, $E^*$) by N(0)=$(1.40, 0.33)$/(eV-f.u.), $v_F$=($5.1, 4.6$)$\times 10^7$cm/s, and $\hbar\Omega_{P}$=$\frac{4\pi e^2}{2}N(0)v_{F}^2$=$(3.8, 1.6)$ eV, quantities required for analysis of data when they become available.

In the isotropic relaxation time approximation, the Hall conductivity $\sigma_{xy}$ involves the integral of the curvature \cite{Ong1991} over each FS. 
For a single circular FS (the $dp\sigma$ FS is also nearly circular; see the SM \cite{sm}), with constant velocity $v$ and curvature $\propto$1/$k_F$ on each surface, each FS contribution is analytic: $\sigma_{xy}=\frac{e^3B}{2\hbar^2}v^2\tau^2$, depending on the square of $\tau$. 
The  Hall coefficient becomes
\begin{eqnarray}
R_H=\frac{\sigma_{xy}} {\sigma_{xx}\sigma_{yy}}
  = \frac{B}{e} \frac{(\tau_p^2 -\tau_n^2)}{[\tau_p p^{1/2} + \tau_n n^{1/2}]^2}.
\end{eqnarray}
The expression can be generalized to accommodate the slightly different carrier velocities if desired. In the limit of large $\tau_n/\tau_p$ this becomes $R_H/B=-1/ne$, i.e., the Hall coefficient becomes dominated by the small $E^*$  ($n$) contribution. Hall measurements on \lnof~will be instructive, potentially verifying the $E^*$ contribution to transport. 

{\it Possible two-gap SC.} 
The evidence above of very weak coupling of $E^*$ to the two most prominent interactions (and mechanisms) promotes a two-band, perhaps extreme two-gap, picture of possible \lnof~SC. In a simple two-gap system, a dominant order parameter (giving $T_c$) is coupled to a subservient order parameter (its $T_c$=0). The BCS treatment of such a two-gap system is simple \cite{Zhitomirsky2004}: A small $E^*$ gap would be induced by the stronger coupling partner, with the ratio of gaps depending on the interband coupling parameter and independent of temperature. A more extensive exposition of Ginzburg-Landau theory \cite{Zhitomirsky2004} should become relevant. This state of well-separated gaps would make specific heat a crucial experiment (as well as others, such as NMR), as well as impacting other SC state parameters.

{\it Summary.}
The properties discussed here pinpoint \lnof~as a unique Ni$^{1+}$ system. Supplemented by a single $E^*$ band of $I$ density spanning all three La layers, it is found that the $I$ density effects span a network of $d$ bands: a $d_{z^2}^e,d_{z^2}^o$ strongly bound pair yet split by 1 eV: the unanticipated $D^*$ pair $d_{xz},d_{yz}$ approaching degeneracy from below, and the $E^*$ band itself; the last two form an incipient degeneracy (Dirac point) that leads to the linearity of the $E^*$ band crossing E$_F$. Whereas the $dp\sigma$ band will show Ni spin fluctuations requiring correlation effects and presenting a short mean free path, the $E^*$ band will likely experience weak renormalization due to only indirect effects from moments and lattice motion, with a longer relaxation time. Effects on electronic transport properties have been outlined.  

While the $E^*$ band provides hole doping to valence Ni$^{1.09+}$, further doping will promote \lnof~into the SC regime, where the $E^*$ band should affect properties in novel ways. Higher doping (higher fraction of anionic F or cationic La$\rightarrow$Sr substitution) and in-plane compressive strain will lead toward the Dirac nonanalyticity of the $D^*$ and $E^*$ bands at the zone corner. The out-of-plane coherence length will likely be on the order of the blocking layer separation, with the next question being whether Josephson coupling between bilayers can promote a coherent 3D SC state. It remains clear, however, that Ni$^{1+}$ is fundamentally different from Cu$^{2+}$.

\vskip 4mm
{\it Acknowledgments}$-$KWL was supported by National Research Foundation of Korea (NRF) Grant (RS-2024-00392493).  YJS was supported by the Deutsche Forschungsgemeinschaft (DFG, German Research Foundation) -- CRC 1487, ``Iron, upgraded!'' -- project number 443703006.

\vskip 4mm
{\it Data availability}$-$The data that support the findings of this article are provided in the main text and supplemental information. Additional data would be available from the authors upon reasonable request.

\bibliography{main}

\vskip 8mm
{\bf End of Paper}

\newpage\onecolumngrid
\renewcommand\thefigure{S\arabic{figure}} % modify \thefigure, *not* \figurename
\newcounter{myfig} % create an extra counter variable
\setcounter{figure}{0}

\renewcommand\thetable{S\arabic{table}} % modify \thefigure, *not* \figurename
\newcounter{mytable} % create an extra counter variable
\setcounter{table}{0}

\renewcommand{\thesection}{S\arabic{section}}
\setcounter{secnumdepth}{1}

\begin{center}
{\bf\Large{Supplemental Material}}
\end{center}

\vskip 10mm
This Supplemental Material provides additional information to provide support to the text. The information is described in the figure captions.

\section{Fermiology} 

\begin{figure*}[htb]
\centering
\includegraphics[width=0.95\columnwidth]{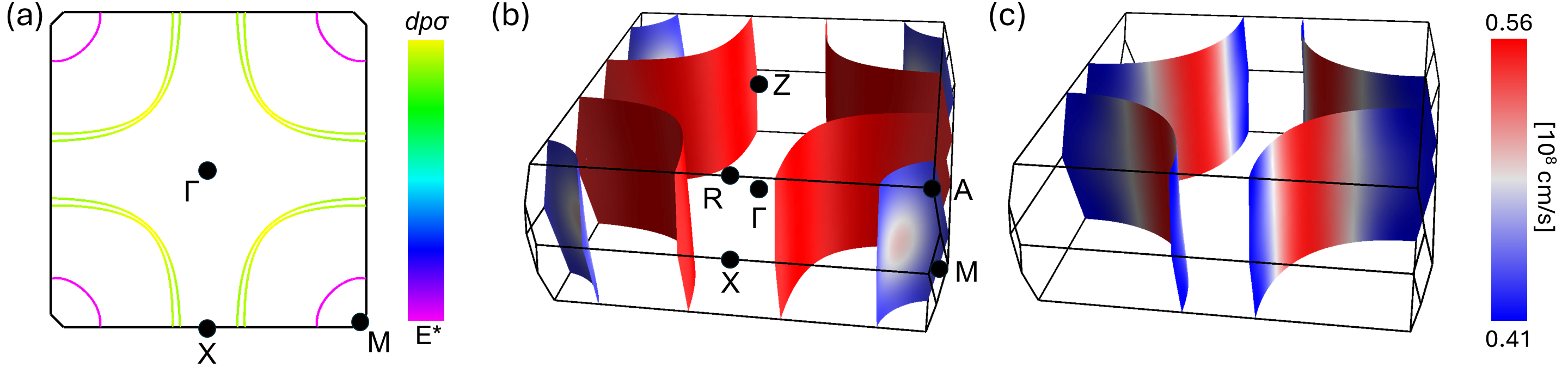}
\caption{ (a) Two-dimensional projected Fermi surfaces (FSs) of nonmagnetic \lnof~within GGA, colored by the orbital characters. The small cylinder arises from the interstitial $E^*$ state.
(b-c) Corresponding three-dimensional FS plots, colored by Fermi velocities. The two large $dp\sigma$ FSs are degenerate except near the zone edge $X$ point, but the velocity varies over one of the FS sheets.}
\label{fig:fs}
\end{figure*}

\clearpage

\begin{figure*}[t]
\centering
\section{Plots of supercell band structure and electron localization function}
\vskip 4mm

\includegraphics[width=0.75\columnwidth]{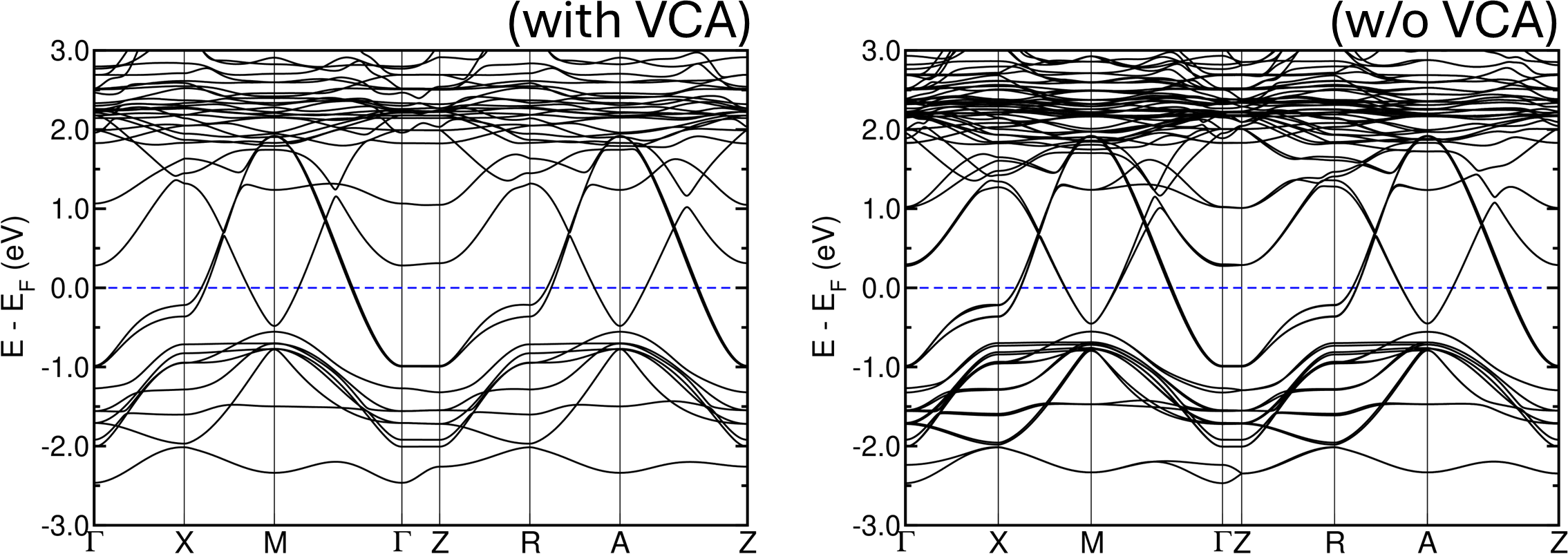}
\caption{Comparison of the band structures of \lnof~obtained using (left) the virtual crystal approximation (VCA) and (right) a supercell approach having twice as many bands (not so obvious).
In the ordered O/F supercell, we placed two O ions and two F ions in the square.
As expected, the two band structures are extremely similar,
since the $2p$ orbitals of these atoms are far below $E_F$ and both ions have filled $2p$ shells.
}
\label{fig:bs}
\end{figure*}

\vskip 4mm

\begin{table*}[tb]
\centering
\caption{Bader charges ($Z$ minus electronic charge) of La$_3$Ni$_2$O$_4$(O$_{1-x}$F$_{x}$)$_2$ at F(0.5) and F(0.700) for comparison. Note that Bader charges do not reproduce the formal charges very closely, as is common. Certain charges change with F content; Ni net charge decreases 0.05$e^-$ with F content. It is worthy of note that the F/O layer X atom does not lose 0.5$e^-$ charge, but only (Bader) 0.109$e^-$ per X ion. Ni layer O and all La charges hardly change.
}
\label{table1}
\begin{tabular}{cccccc}\hline\hline
       ~~~atoms~~~& ~~~ multiplicity~~~&~~~~~~&  ~~F(0.5)~~&~~~~~~& ~~F(0.700)~~\\ \hline
       ~~~Ni~~~&  2&~~~&  ~~0.556 &~&   ~~0.501 \\ 
       O&  4&~~~&  $-1.244$&~&   $-1.247$\\ 
       F/O&  2&~~~&  $-1.118$&~& $-1.028$\\  
       La1&  1&~~~&  ~~1.938&~&   ~~1.924\\  
       La2&  2&~~~&  ~~2.087&~&   ~~2.062\\  \hline
      sum&  -- &~~~&  ~~0.010&~&   ~~0.001\\  \hline \hline
\end{tabular}
\end{table*}

\vskip 4mm

\begin{figure*}[t]
\centering
\includegraphics[width=0.75\columnwidth]{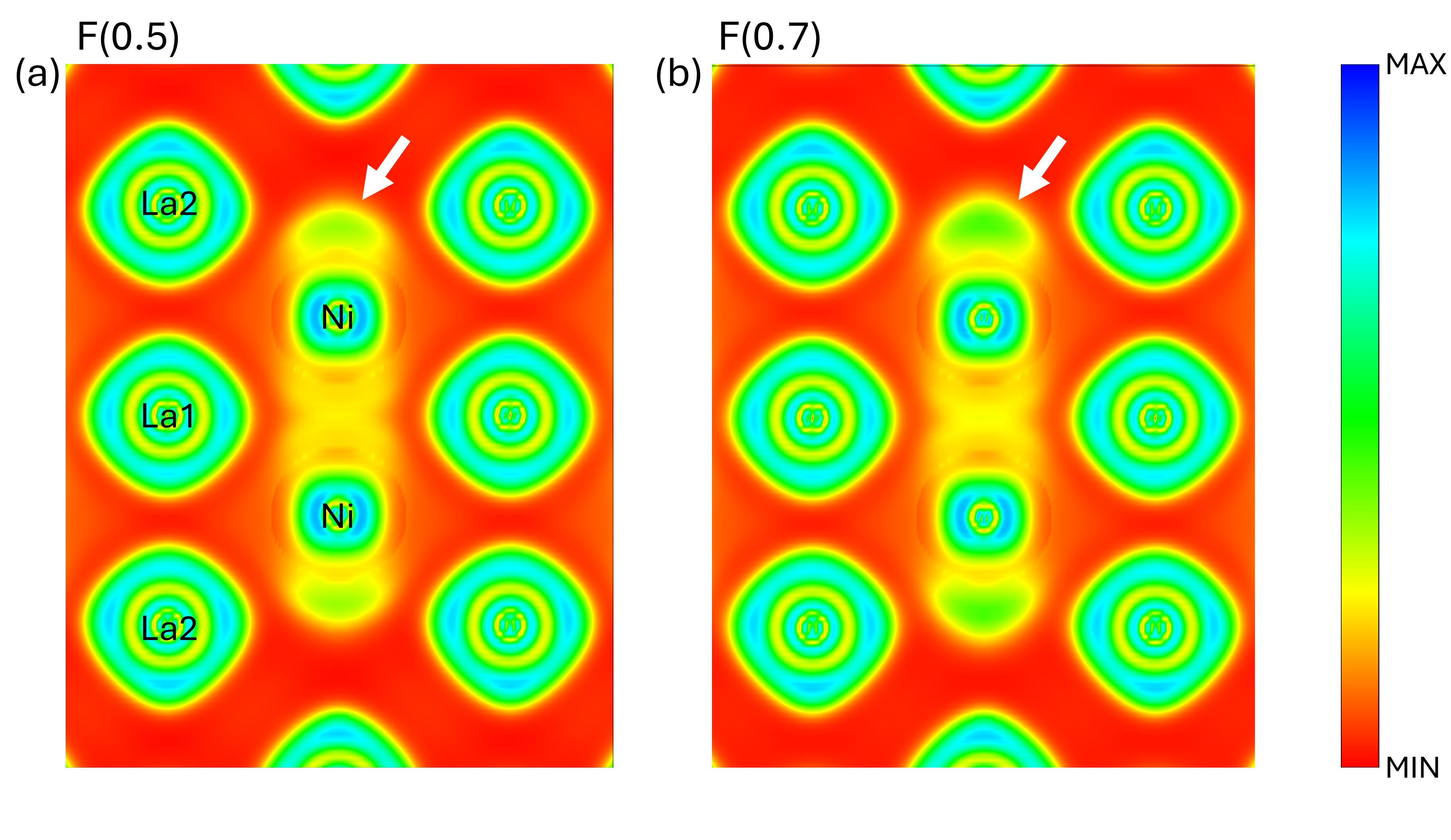}
\caption{Electron localization function plots on the [110] plane at (a) F(0.5) and (b) F(0.7). The white arrow indicates a density maxium between the NiO$_2$ and La2 layers, and double-crescent-shaped weaker local maxima in the La1 layer. These maxima become somewhat more prominent with increasing F content. These results are roughly consistent with the plots of the $E^*$ wavefunction.
}
\label{fig:elf}
\end{figure*}

\clearpage
\section{Symmetry Analysis and Isosurface Contours}

\begin{figure*}[h]
\centering
\includegraphics[width=1.0\columnwidth]{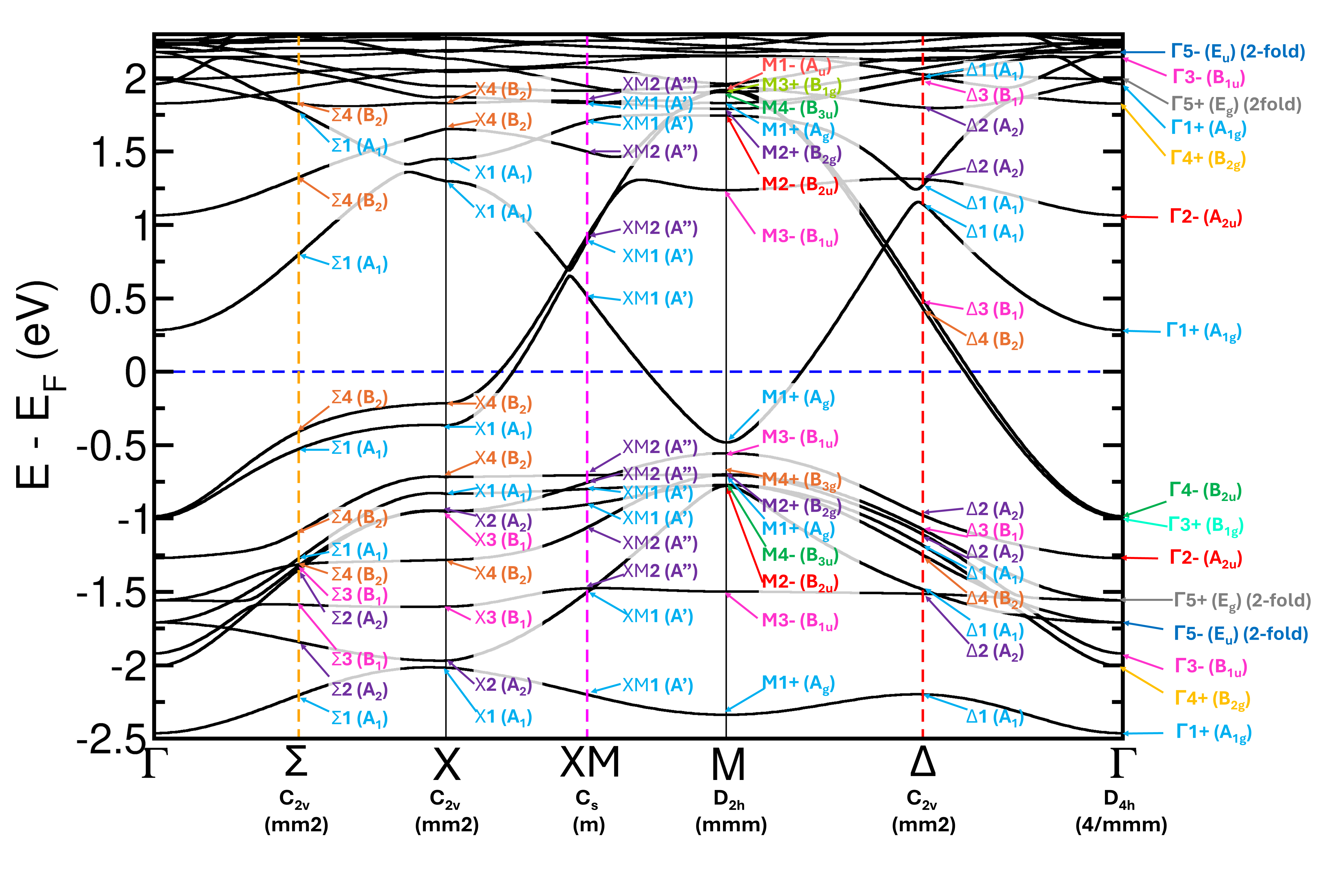}
\caption{GGA band structure of \lnof~(the $X$ $2p$ bands lie lower), providing the irreducible representations of each band at several high symmetry points in the $I4/mmm$ space group.  The $E^*$ band dips well below the Fermi level (the zero of energy) at $M$.
}
\label{fig:nm_ir_wf}
\end{figure*}

\begin{figure*}[h]
\centering
\includegraphics[width=0.9\columnwidth]{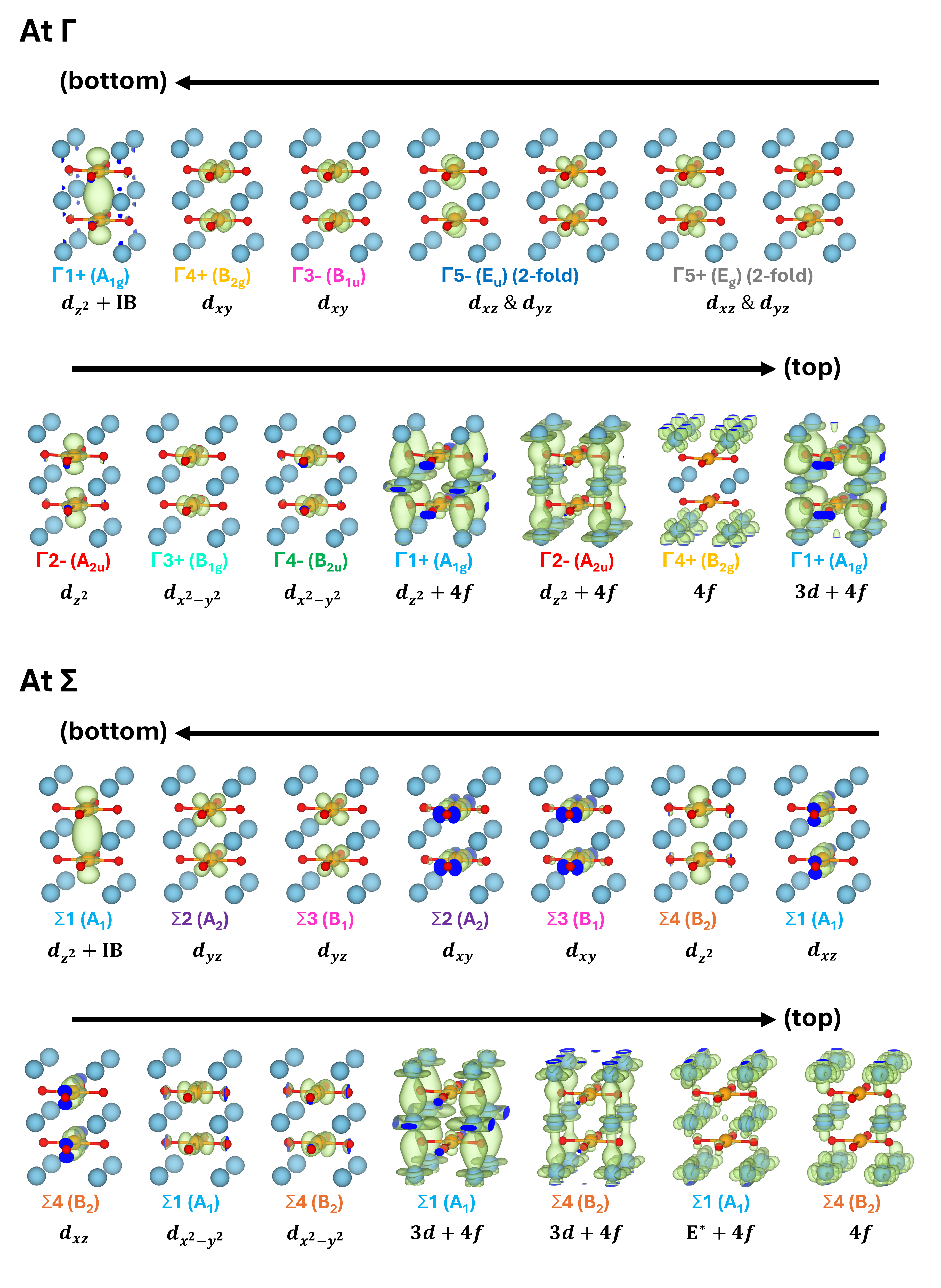}
\caption{Isosurface plots of wave functions $|\Psi(r)|$ for each eigenstate from the bottom to the top at the $\Gamma$ and $\Sigma$ points. Extended lime colored regions indicate contributions from the interstitial density. This is the case for several following plots. The upper labels indicate the symmetry, while the lower labels give a description of the character of the state.}
\label{fig:nm_wf1}
\end{figure*}

\begin{figure*}[h]
\centering
\includegraphics[width=0.9\columnwidth]{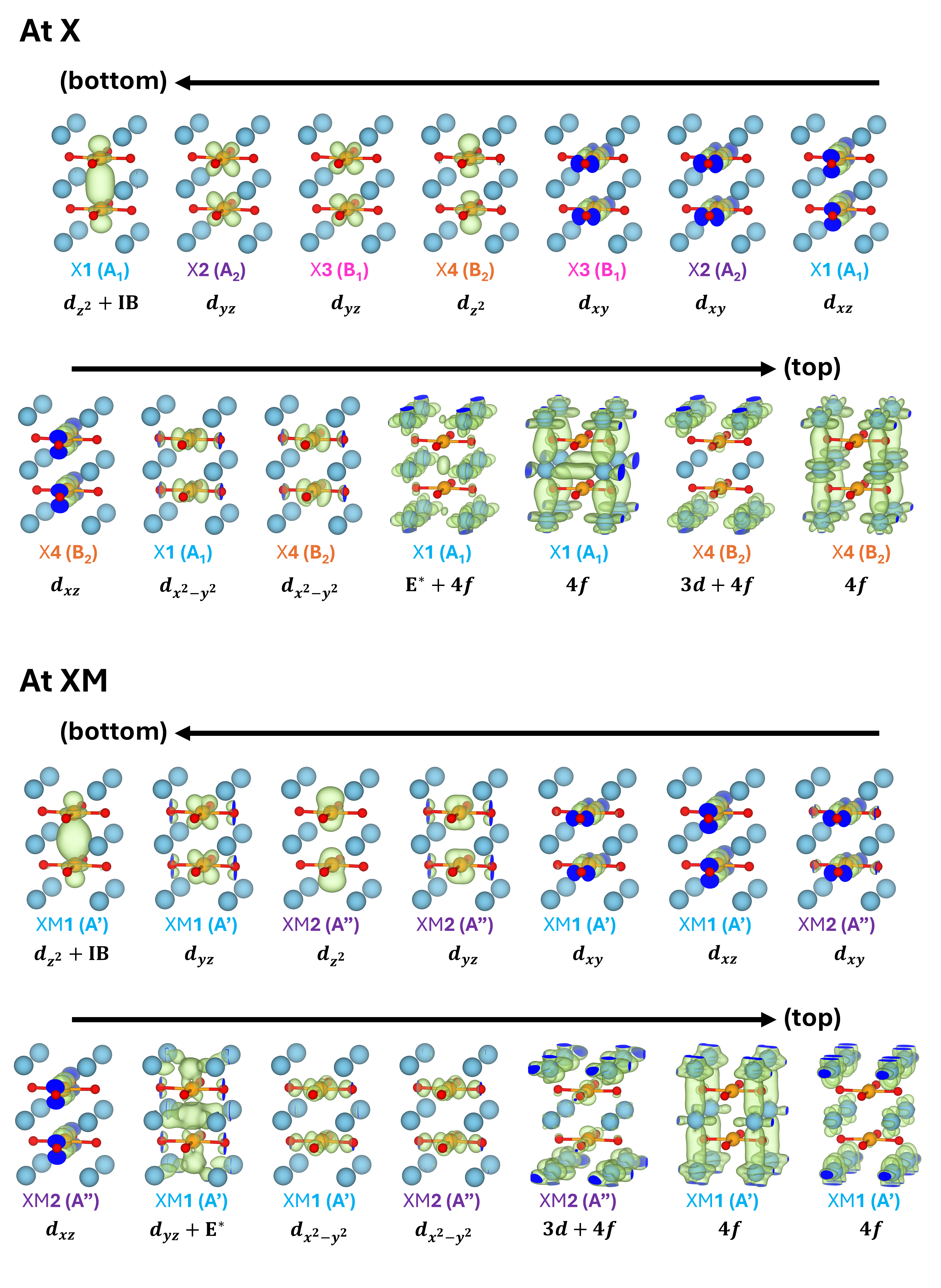}
\caption{Isosurface plots of wave function plots for each point from the bottom to the top at the $X$ and $XM$ points, with labeling as above.}
\label{fig:nm_wf2}
\end{figure*}

\begin{figure*}[h]
\centering
\includegraphics[width=0.9\columnwidth]{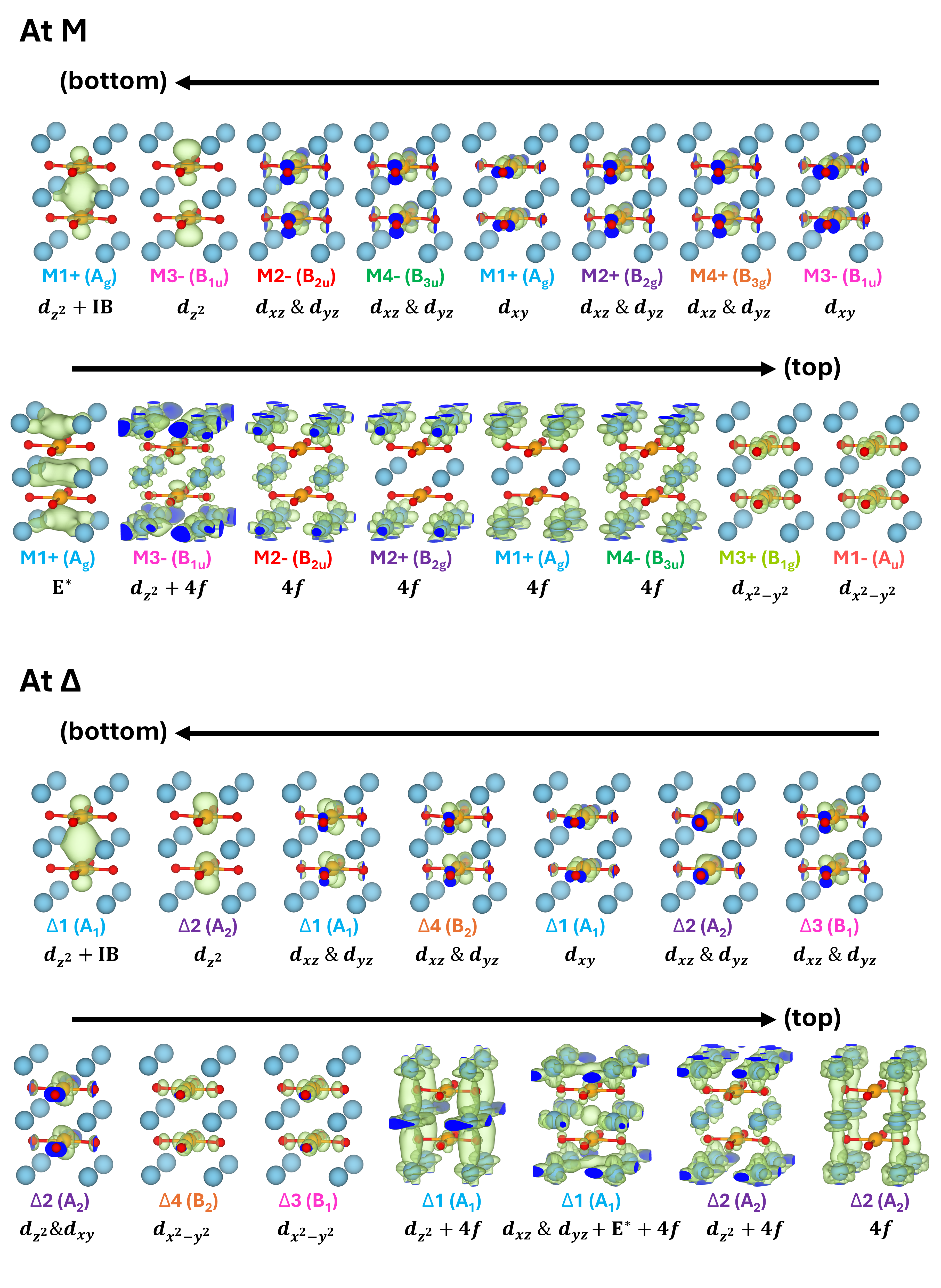}
\caption{Isosurface plots of wave functions for each eigenvalue from the bottom to the top at the $M$ and $\Delta$ points. 
}
\label{fig:nm_wf3}
\end{figure*}

\clearpage
\section{Variations of $D^*$ and $E^*$ bands by F concentration, B\lowercase{a}-doping, phonons, lattice parameter, and pressure}

\vskip 8mm
\begin{figure*}[h]
\centering
\includegraphics[width=1.0\columnwidth]{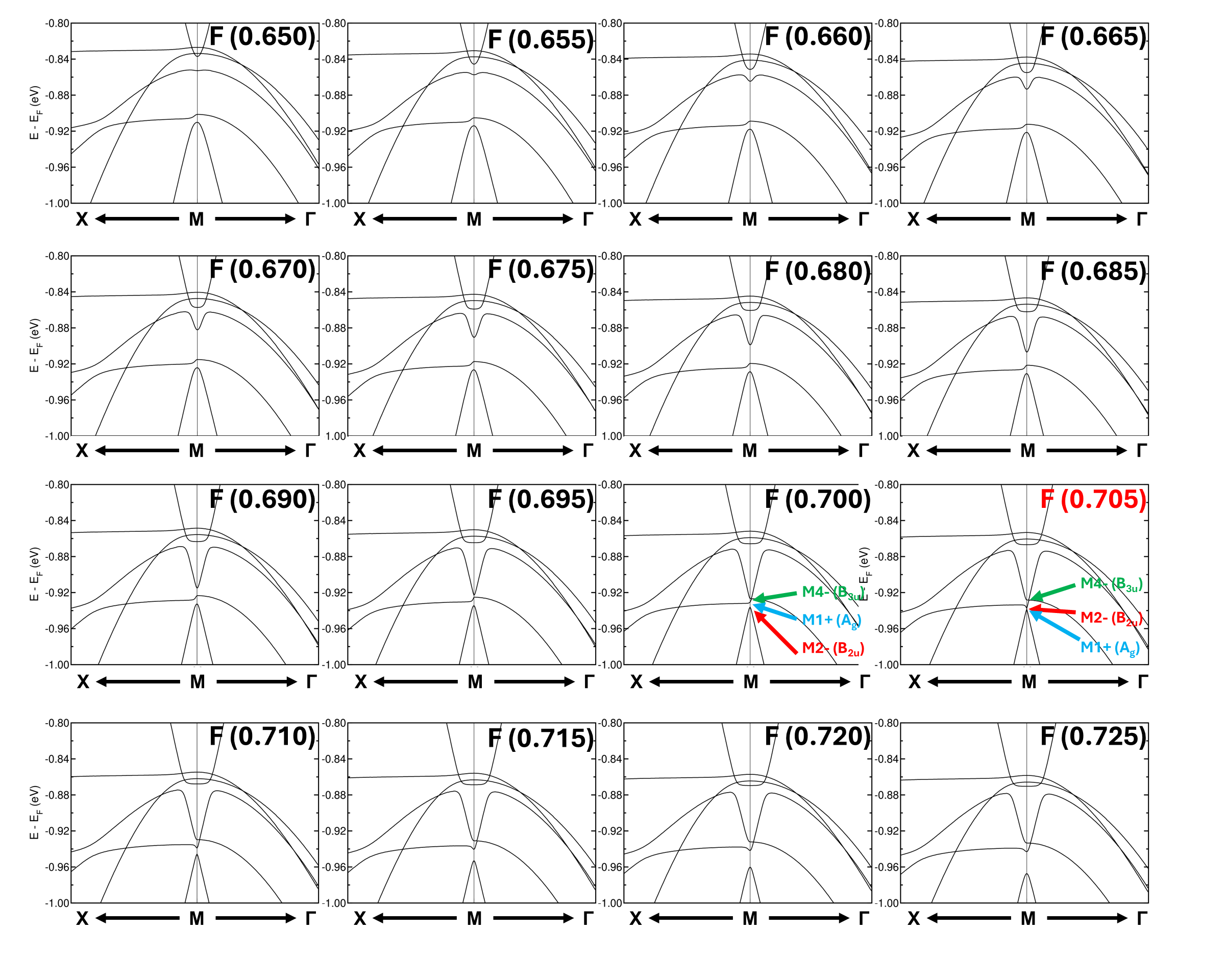}
\caption{Evolution of the $E^*$ band and related bands versus F occupancy from 0.650 to 0.725 in La$_3$Ni$_2$O$_4$(O$_{1-x}$F$_{x}$)$_2$. At the critical doping level of 0.705,  
corresponding to La$_3$Ni$_2$O$_{4.59}$F$_{1.41}$, the $E^*$ band [$M_1^+$ ($A_g$)] forms a Dirac point at around $-0.94$ eV with the odd-parity $d_{xz}$,$d_{yz}$ $D^*$ band [$M_2^-$ ($B_{2u}$)]. Upon further doping, the crossing lifts and the bands separate.}
\label{fig:nm_dp_evo}
\end{figure*}

\begin{figure*}[htb]
\centering
\includegraphics[width=0.9\columnwidth]{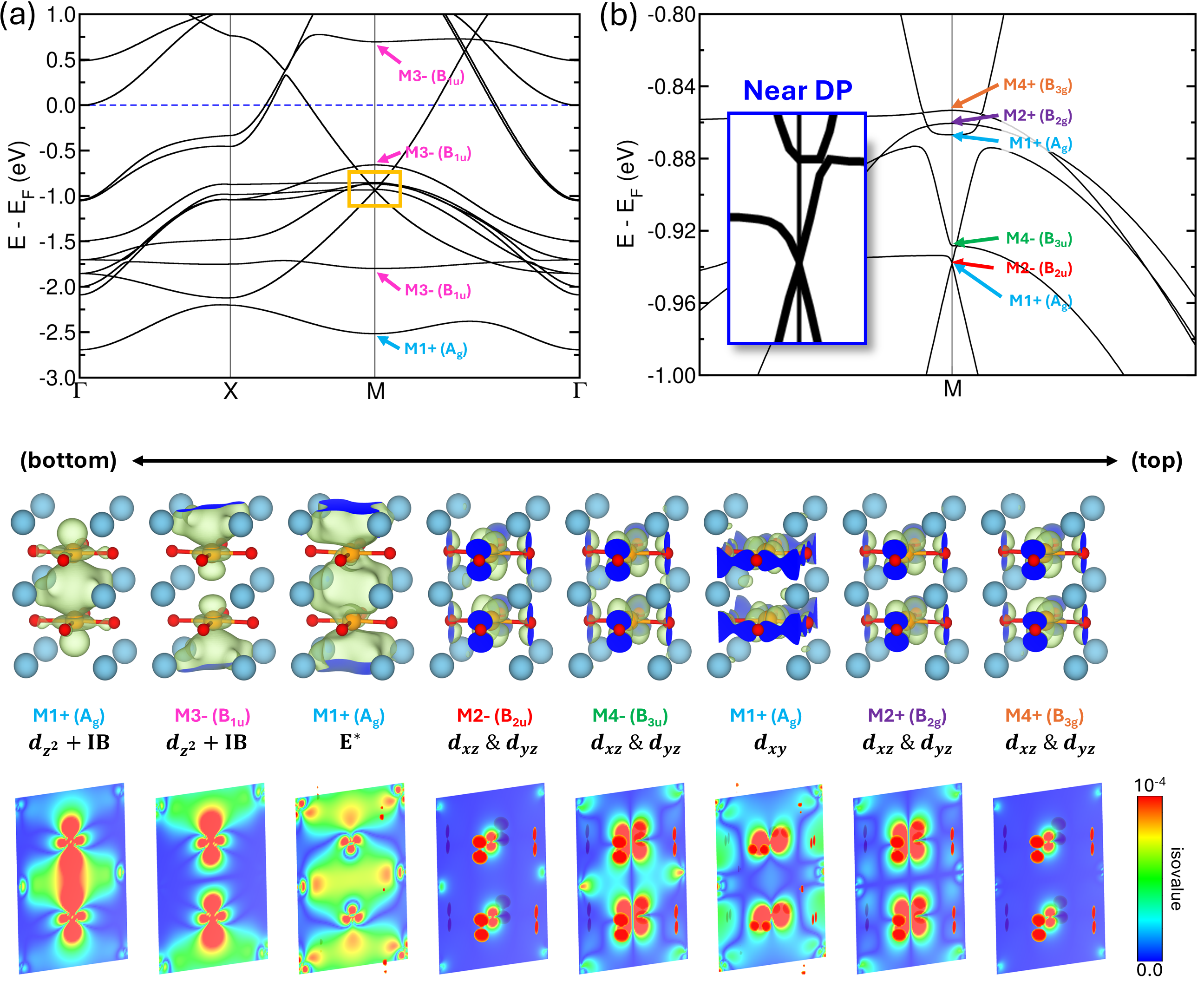}
\caption{(a) Band structure of La$_3$Ni$_2$O$_{4.59}$F$_{1.41}$, {\it i.e.} at the F fraction of 0.705, together with the irreducible labels at the $M$ point where the singular point occurs. (b) The corresponding band structure is zoomed in (twice) near the linear crossing point. At the bottom are a series of wave-function isosurface plots of the wavefunctions at the $M$ point, from the bottom energy up, together with their 2D-projected isocontour maps on the [110] plane. The first two are those designated in the text $d_{z^2}^e,d_{z^2}^o$ (even, odd), which illustrates the participation of the interstitial density; they are very much the same for F(0.5). The third ($E^*$) eigenvalue (blue $M_1^+$), with the interstitial density, has crossed the $D^*$ ($M_4^-$) state by a tiny amount.  The involvement of the $M_4^-$ density with interstitial density, though relatively minor, identifies it as the $D^*$ partner of $E^*$.
}
\label{fig:nm_f70}
\end{figure*}

\begin{figure*}[htb]
\centering
\includegraphics[width=0.98\columnwidth]{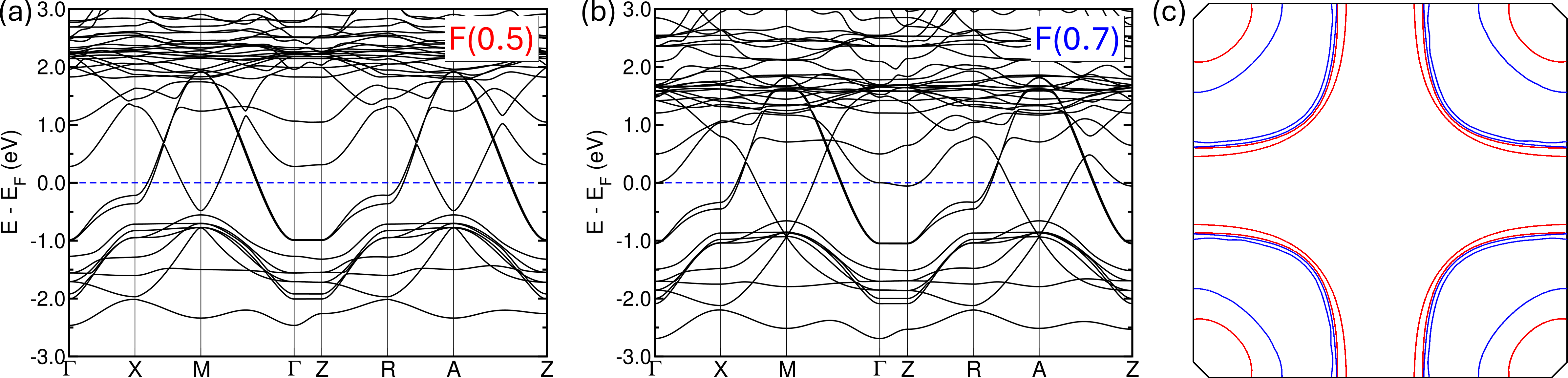}
\caption{Variations in the size of the $E^*$ electron FS with changing F concentration in La$_3$Ni$_2$O$_4$(O$_{1-x}$F$_{x}$)$_2$. (a), (b) show the band structures at F(0.5) and F(0.7), respectively, with (c) the corresponding FSs. 
Increasing the F ratio from 0.5 to 0.7 leads to a some enlargement of the $E^*$ electron pocket and electron density. The $dp\sigma$ band thereby acquires more additional holes than the F difference (0.2) provides. Specifically, the $E^*$ electron pocket has a 45\% larger radius $k_F$ at F(0.7) than at F(0.5), so the Ni valence is +1.19, just below the dome peak.
}
\label{fig:S7b}
\end{figure*}

\begin{figure*}[h]
\centering
\includegraphics[width=1.0\columnwidth]{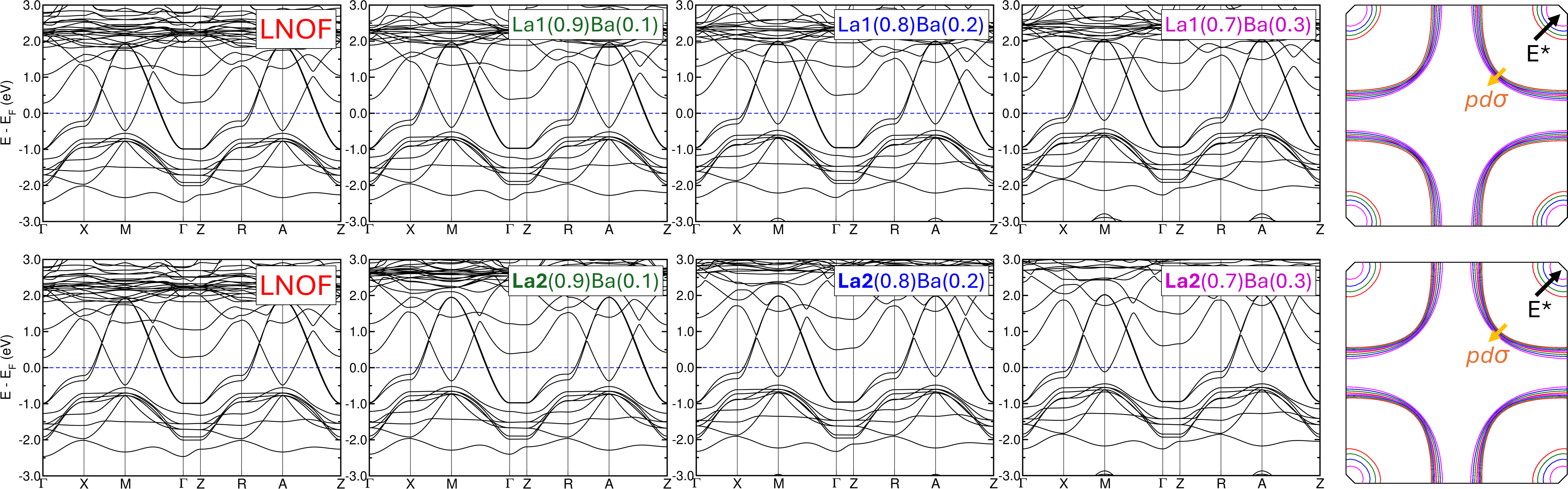}
\caption{Band structures of (La,Ba)$_3$Ni$_2$O$_5$F, at different Ba doping levels from 0.1 to 0.3, together with (rightmost panels) the variations of the FSs.
Note that the VCA approach is applied to either (upper row) La1 or (lower row) La2. The main change is in the self-doping, which is very much as would be expected from a rigid band approximation. At 30\% Ba-doping level, the electron carriers contained in the $E^*$ FSs reduce to 0.072$e^-$ and 0.052$e^-$ for the La1 and La2 cases, respectively. The result is that doping of holes by the La to Ba substitution negates the intrinsic electron doping of \lnof, as would be expected from usual carrier counting.
}
\label{fig:ba-doping}
\end{figure*}

\begin{figure*}[h]
\centering
\includegraphics[width=1.0\columnwidth]{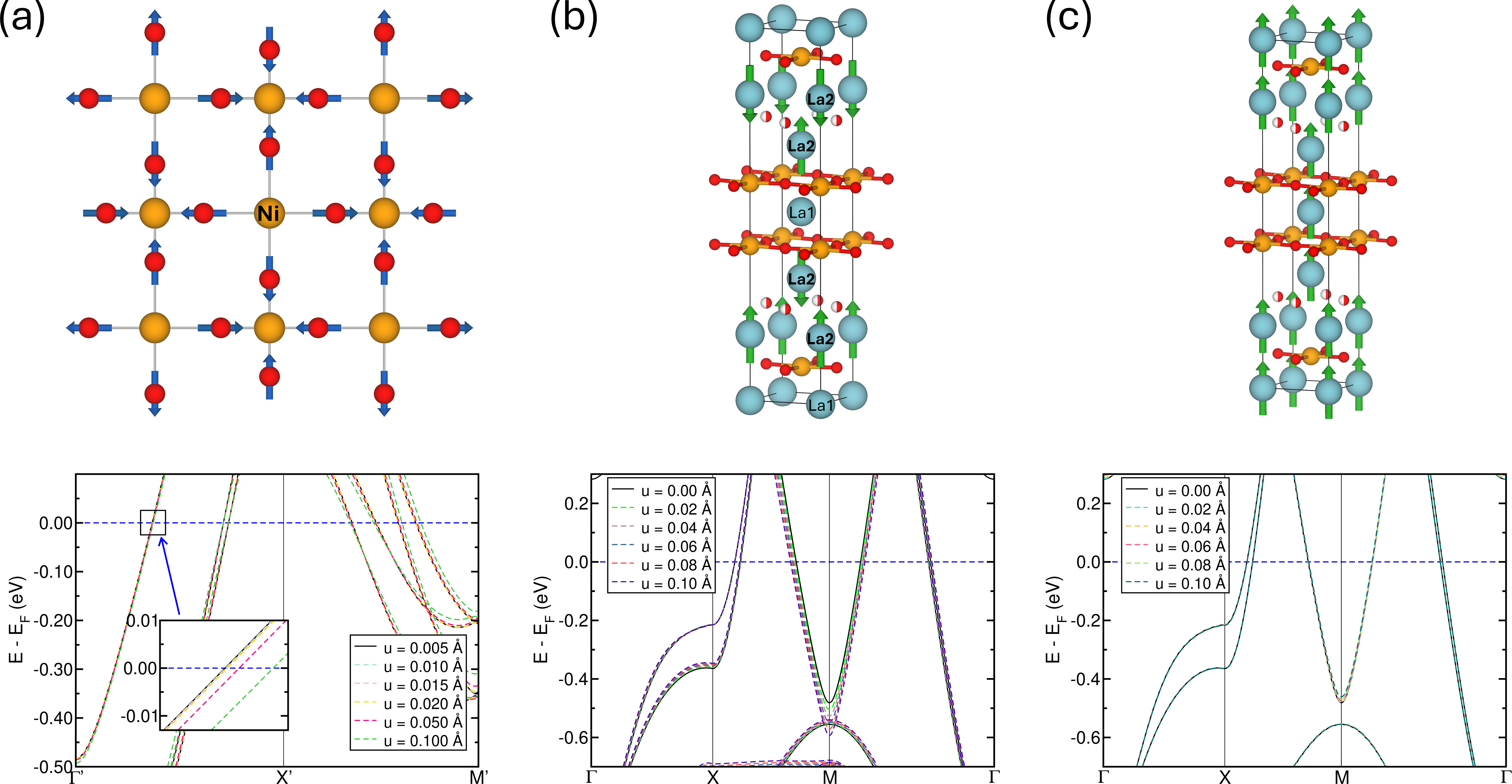}
\caption{Frozen phonon calculations for the oxygen breathing mode in the NiO$_2$ layers and vertical vibrating modes of La ions, in \lnof.
Top: Sketches of ion vibrations for (a) the oxygen breathing mode, (b) La2 $A_{1g}$, and (c) La $A_{2u}$ modes.
Bottom: Corresponding variations of bands at $E_F$ for each mode.
These yield tiny deformation potentials of 0.1--1 eV/\AA~for the E* band,
indicating minor phonon effects.
}
\label{fig:S7c}
\end{figure*}

\begin{figure*}[h]
\centering
\includegraphics[width=1.0\columnwidth]{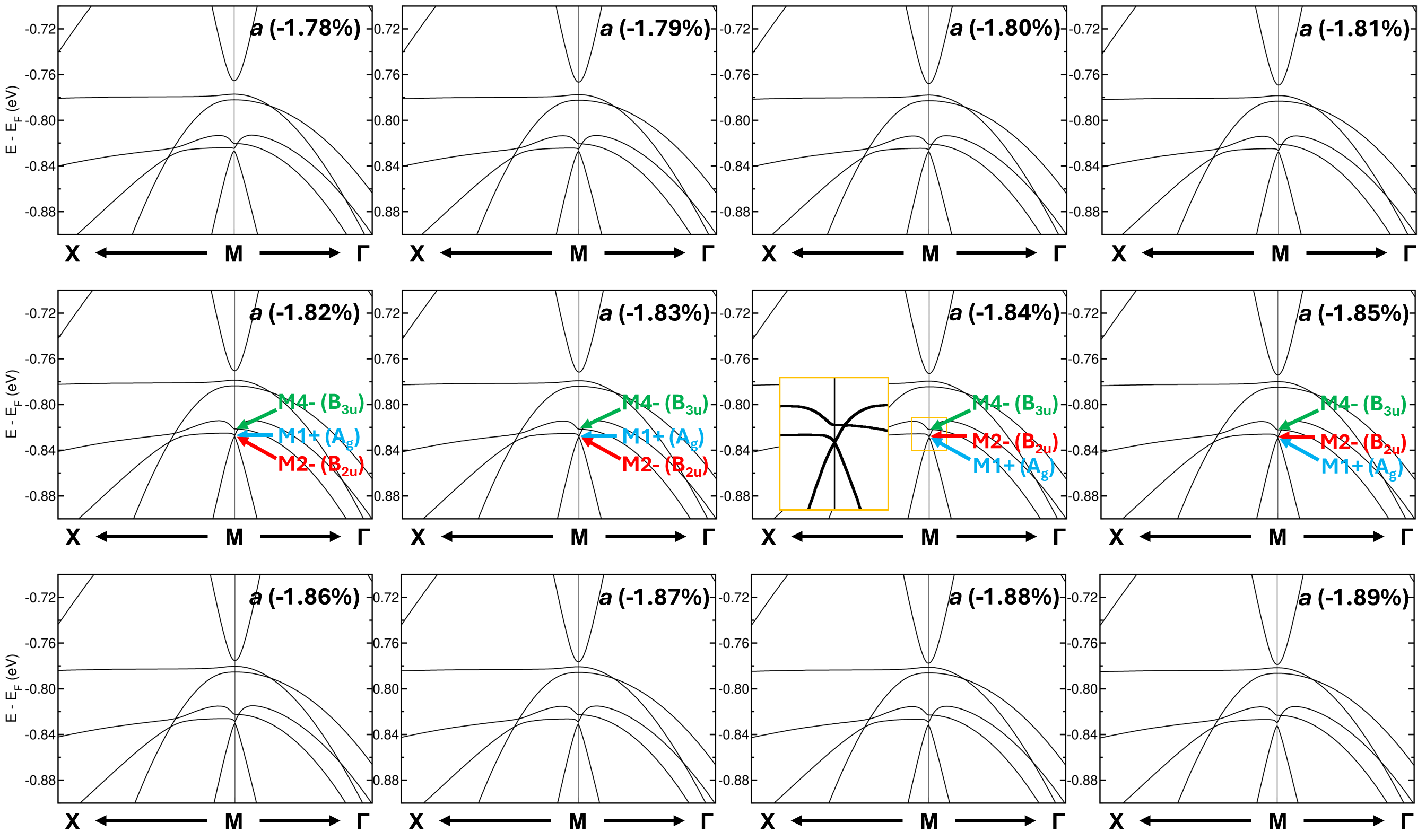}
\caption{Evolution of the $E^*$ band with variation of the $a$ lattice constant, fine-tuned from $-1.78$\% to $-1.89$\%, with fixed volume in \lnof. The multiband complex complicates the behavior, which is quite sensitive to compression, but the Dirac point degeneracy occurs very close to the $-1.84$\% compression. It is interesting that the ``initial'' and ``final'' band structures look virtually the same, in spite of the changing of the order of states. A larger scale variation is provided in the next figure.
}
\label{fig:nm_dp_evo3}
\end{figure*}

\begin{figure*}[h]
\centering
\includegraphics[width=1.0\columnwidth]{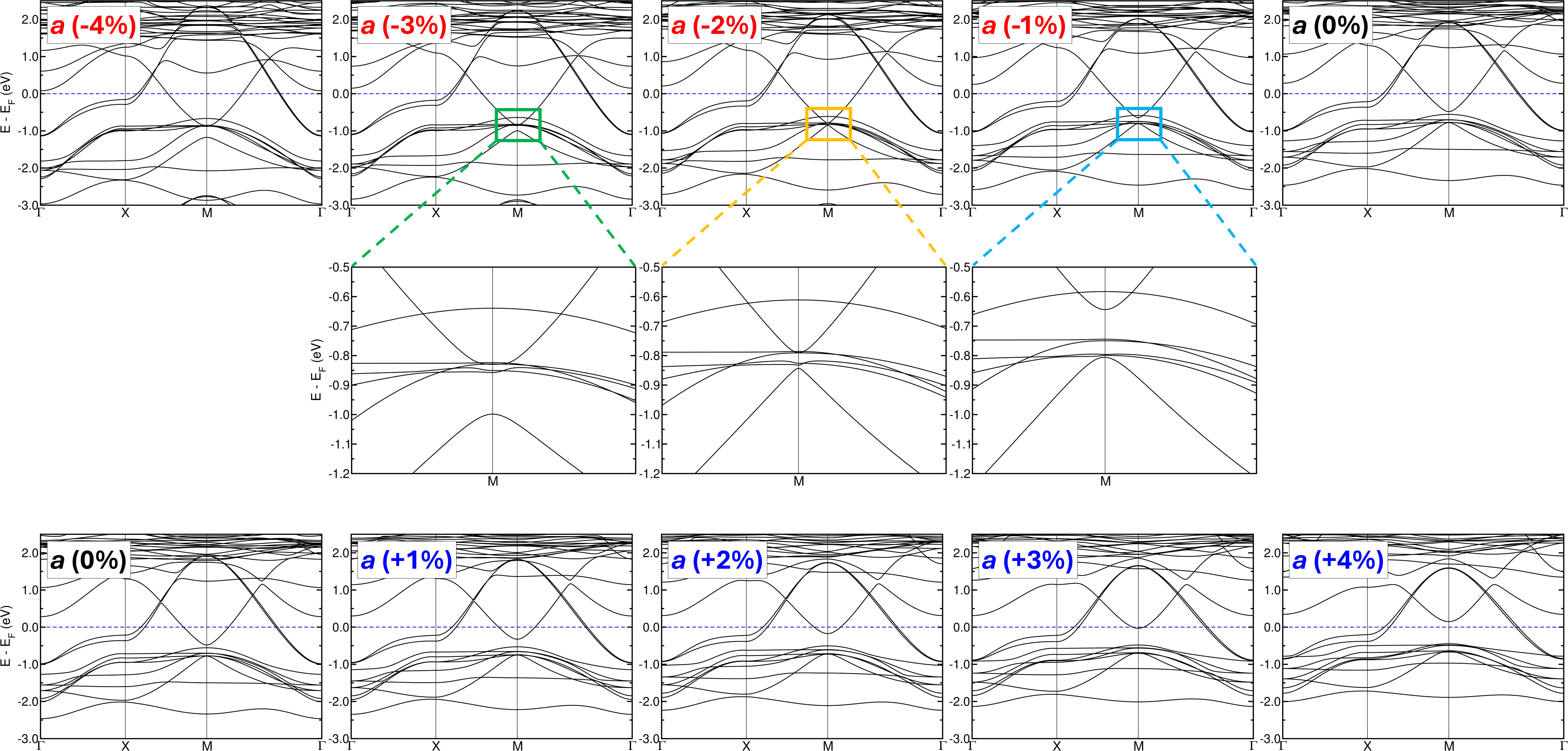}
\caption{Evolution of the $E^*$ band position with variation of the $a$ lattice constant, broad-tuned from $-4$\% to $+4$\%, with fixed volume in \lnof. In addition to the larger scale view of variation of self-doping, the figure provides a close-up picture of the band complex very near ($-2$\%) the critical Dirac point. 
}
\label{fig:nm_dp_evo4}
\end{figure*}

\begin{figure*}[h]
\centering
\includegraphics[width=0.99\columnwidth]{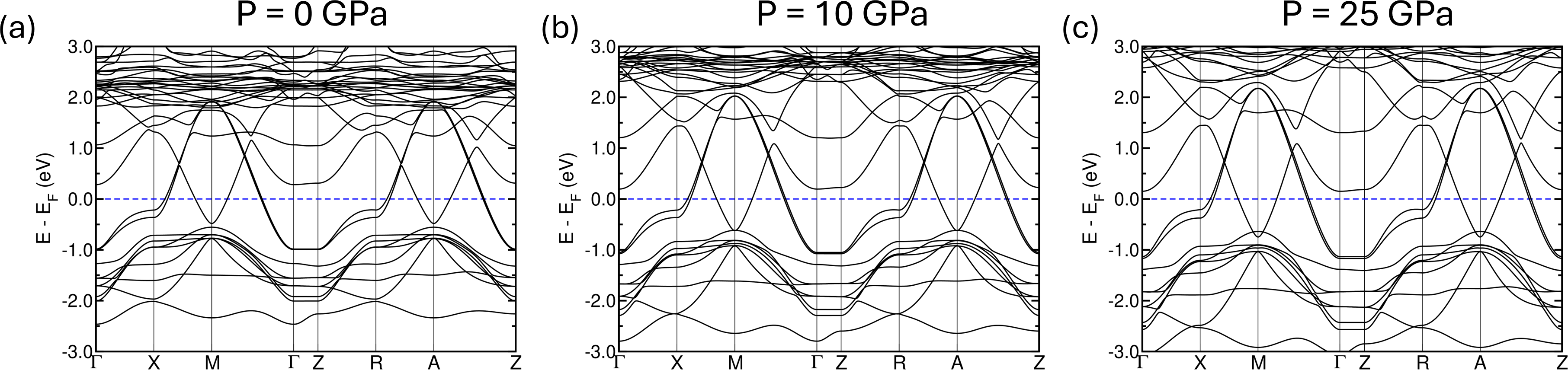}
\caption{Variations of band structure of \lnof~at hydrostatic pressure $P$ of (a) 0 GPa, (b) 10 GPa, and (c) 25 GPa.
By applying external hydrostatic pressure, the $E^*$ and $D^*$ FSs increase in area (density). Fine-tuning to somewhat higher $P$ values will produce the Dirac crossing as does the simple reduction of the $a$ lattice parameter.
}
\label{fig:bs_F-ratio}
\end{figure*}

\end{document}